\documentstyle[12pt]{article}

\newcommand{\mpl}{M_{\rm pl}}
\newcommand{\mst}{M_{*}}

\newcommand{\be}{\begin{equation}}
\newcommand{\ee}{\end{equation}}
\newcommand{\bea}{\begin{eqnarray}}
\newcommand{\eea}{\end{eqnarray}}

\newcommand{\newc}{\newcommand}
\newc{\gsim}{\lower.7ex\hbox{$\;\stackrel{\textstyle>}{\sim}\;$}}
\newc{\lsim}{\lower.7ex\hbox{$\;\stackrel{\textstyle<}{\sim}\;$}}
\newc{\gev}{\,{\rm GeV}}
\newc{\mev}{\,{\rm MeV}}
\newc{\ev}{\,{\rm eV}}
\newc{\kev}{\,{\rm keV}}
\newc{\tev}{\,{\rm TeV}}

\renewcommand{\phi}{\varphi}
\newc{\smu}{{\tilde\mu}}
\newc{\snu}{{\tilde\nu}}
\newc\order{{\cal O}}
\newc{\eps}{\epsilon}
\newc{\re}{\mbox{Re}\,}
\newc{\im}{\mbox{Im}\,}
\newc{\lunits}{\,\mbox{cm}^{-2}\mbox{s}^{-1}}
\relax
\def\boxeqn#1{\vcenter{\vbox{\hrule\hbox{\vrule\kern3pt\vbox{\kern3pt
\hbox{${\displaystyle #1}$}\kern3pt}\kern3pt\vrule}\hrule}}}
\def\qed#1#2{\vcenter{\hrule \hbox{\vrule height#2in
\kern#1in \vrule} \hrule}}
\newc{\ie}{{\it i.e.}} \newc{\etal}{{\it et al.}}
\newc{\eg}{{\it e.g.}} \newc{\etc}{{\it etc.}}
\newc{\cf}{{\it c.f.}}
\def\ltap{\ \raise.3ex\hbox{$<$\kern-.75em\lower1ex\hbox{$\sim$}}\ }
\def\gtap{\ \raise.3ex\hbox{$>$\kern-.75em\lower1ex\hbox{$\sim$}}\ }
\def\gl{\ \raise.5ex\hbox{$>$}\kern-.8em\lower.5ex\hbox{$<$}\ }
\def\roughly#1{\raise.3ex\hbox{$#1$\kern-.75em\lower1ex\hbox{$\sim$}}}
\def\dsl{\,\raise.15ex\hbox{/}\mkern-13.5mu D} 
\def\delsl{\raise.15ex\hbox{/}\kern-.57em\partial}
\def\Ksl{\hbox{/\kern-.6000em\rm K}}
\def\Asl{\hbox{/\kern-.6500em \rm A}}
\def\Dsl{\hbox{/\kern-.6000em\rm D}} 
\def\Qsl{\hbox{/\kern-.6000em\rm Q}}
\def\gradsl{\hbox{/\kern-.6500em$\nabla$}}
\let\beta=\beta

\let\ka=\kappa

\newdimen\pmboffset
\def\oldpmb#1{\setbox0=\hbox{#1}%
\copy0\kern-\wd0
\kern\pmboffset\raise 1.732\pmboffset\copy0\kern-\wd0
\kern\pmboffset\box0}

\def\bar#1{\overline{#1}}

\def\inv{^{\raise.15ex\hbox{${\scriptscriptstyle -}$}\kern-.05em 1}}
\def\pr#1{#1^\prime} 
\def\lbar{{\lower.35ex\hbox{$\mathchar'26$}\mkern-10mu\lambda}} 

\let\<=\langle
\let\>=\rangle

\let\+=\uparrow
\let\-=\downarrow

\begin{document}

\baselineskip=17pt
\pagestyle{plain}
\setcounter{page}{1}

\begin{titlepage}

\begin{flushright}
SLAC-PUB-8014\\
SU-ITP-98/64

\end{flushright}
\vspace{5 mm}

\begin{center}
{\LARGE Neutrino Masses from Large Extra Dimensions \footnote{A preliminary version of this work was presented
at SUSY 98, in July, 1998 \cite{1}.}}
\vskip 2mm
\end{center}

\begin{center}
{\large Nima Arkani-Hamed$^a$, Savas Dimopoulos$^b$, Gia Dvali$^c$
\\ and John March-Russell$^d$\\
\vspace{2mm}
{\em $^a$ SLAC, Stanford University, Stanford CA 94309,
USA}\\
{\em $^b$ Physics Department, Stanford University, Stanford CA 94305,
USA}\\
{\em $^c$ Physics Department, New York University, NY, 10003, USA \\
ICTP, Trieste, Italy\\}
{\em $^d$ Theory Division, CERN, CH-1211, Geneva 23,
Switzerland}\\}
\end{center}
\vspace{2mm}
\begin{center}
{\large Abstract}
\end{center}
\noindent
Recently it was proposed that the standard model (SM) degrees of
freedom reside on a $(3+1)$-dimensional wall or ``3-brane''
embedded in a higher-dimensional spacetime. Furthermore,
in this picture it is possible for the fundamental Planck mass $\mst$
to be as small as the weak scale $\mst\simeq O(\tev)$ and
the observed weakness of gravity at long
distances is due the existence of new sub-millimeter spatial
dimensions. We show that in this picture it is natural to
expect neutrino masses to occur in the $10^{-1} - 10^{-4}\ev$ range,
despite the lack of any fundamental scale higher than $\mst$. Such
suppressed neutrino masses are not the result of a see-saw,
but have intrinsically higher-dimensional explanations.
We explore two possibilities. The first mechanism identifies
any massless bulk fermions as right-handed
neutrinos. These give naturally
small Dirac masses for the same reason that gravity is weak at long distances in this framework.
The second mechanism takes advantage of the large {\it infrared} desert: the space in the extra dimensions.
Here, small Majorana neutrino masses are generated by breaking lepton
number on distant branes.
\end{titlepage}
\newpage

\section{Introduction}

It has recently become clear that the fundamental scale of gravity
need not be the Planck scale $\mpl\simeq 1.2\times 10^{19}\gev$, but
rather that the true scale $\mst$ where gravity becomes strong
can be much lower.
The observed small value of Newton's constant at long distances
is ascribed to the spreading of the gravitational force in $n$
``large'' extra dimensions. The volume
$R^n$ of the new dimensions is fixed by Gauss' law to be
\be
R^n \simeq \mpl^2/\mst^{n+2}.
\ee
The most radical, and in many ways the most attractive suggestion
for $\mst$, is that it should be close to the weak scale $\mst\sim
1\tev$.
In this case we have
$R \simeq 10^{-17+\frac{30}{n}} \mbox{cm}$.
For $n=1$, $R \sim 10^{13}~{\rm cm}$, so this case is
excluded since it would modify Newtonian gravitation at
solar-system distances. Already for $n=2$, however, $R \sim 1~{\rm mm}$,
which happens to be the distance where our present experimental
knowledge of gravitational strength forces ends.

While the gravitational force has not been measured beneath
a millimeter, the success of the SM up to $\sim 100\gev$ implies
that the SM fields can not feel the extra large dimensions; that
is, they must be stuck on a 3-dimensional wall, or ``3-brane", in the
higher dimensional space. Thus, in this framework the universe is
$(4+n)$-dimensional with fundamental Planck scale $\mst$ residing
somewhere between the weak scale and $\mpl$, with
new sub-mm sized dimensions where gravity,
and perhaps other fields, can freely propagate, but where the SM
particles are localized on a 3-brane in the higher-dimensional space
\cite{ADD,AADD,ADDlong}.

The most attractive possibility for localizing the SM fields to
the brane is to employ the D-branes that naturally occur in type I
or type II string theory \cite{Dbrane,AADD}. Gauge and other
degrees of freedom are naturally confined to such
D-branes~\cite{Dbrane}, and furthermore this approach has the
advantage of being formulated within a consistent theory of gravity.
However, from a practical point of view, the most important question
is whether this framework is experimentally excluded. This was the
subject of \cite{ADDlong} where laboratory, astrophysical, and
cosmological constraints were studied and found not to exclude these
ideas, even for $\mst$ as low as $1\tev$. There are a number of
model independent predictions of such a scenario, ranging from the
production of Regge excitations and bulk gravitons at the next
generation
of colliders \cite{ADD,AADD,recent},
to the modification of the properties of black holes
\cite{ADMR}.

There are also a number of other papers discussing related suggestions.
Refs.~\cite{DDG} examine the idea of lowering the GUT scale by
utilizing higher dimensions. Further papers
concern themselves with the construction of string models with extra
dimensions larger than the string scale \cite{antoniadis,HW,tye},
and gauge coupling unification in higher dimensions
without lowering the unification scale~\cite{bachas}.
There are also important papers by Sundrum on the effective theory of
the low energy degrees
of freedom in realizations of our world as a brane, and on
radius stabilization~\cite{Raman1,Raman2}. For earlier works on the world as a three-dimensional wall,
see \cite{early}. The issue of radius
stabilization was also considered in \cite{AHDMR}.

However, it may seem that we have given up any hope of explaining
the size of the neutrino masses deduced to be necessary
to explain the atmospheric \cite{SuperK} and solar \cite{Solar} neutrino anomalies.
In the traditional approach the small neutrino masses are
the result of the see-saw mechanism, in which a large right-handed
(rhd) majorana mass $M_R$ suppresses one of the eigenvalues of the
neutrino mass matrix, leading to $m_\nu \sim m_{\rm fermion}^2/M_R$.
The neutrino mixing explanations of the atmospheric and solar neutrino
anomalies require $M_R$ to be a superheavy mass scale, varying between
an intermediate scale $\sim 10^{10}\gev$ the GUT scale. However,
in the world-as-a-brane picture with $\mst\sim 1\tev$ the existence of
such a superheavy scale is unpalatable.

In this letter we show that there are {\it intrinsically
higher-dimensional}
explanations for either Dirac or Majorana neutrino masses.
For Dirac masses,
The basic idea is that {\it any} fermionic state that propagates
in the bulk must, by definition, be a SM singlet, and furthermore
that it couples to the wall-localized SM states precisely as a
right-handed
neutrino
with a {\it naturally small coupling}. The small coupling is a result
of the
large relative volume of the internal ``bulk'' manifold compared to the
thin wall where SM states propagate. The interaction probability of the
Kaluza-Klein (KK)
zero mode of the bulk rhd neutrino state $\nu_R$ with the
brane-localized Higgs
and Lepton doublet fields is thus small, resulting in a greatly
suppressed
$\nu_R(x,y=0)L(x) H(x)$ coupling. Small Majorana masses can be
obtained using the generic mechanism of \cite{AHD} for generating
small couplings by breaking symmetries on distant branes in the
bulk. In our context, we break lepton number on far-away branes,
and have this breaking communicated to us by bulk messenger
fields, giving a naturally distance-suppressed Majorana neutrino
mass on our wall.

\section{Right-handed neutrinos in the bulk}

In this section, we will show that neutrinos can acquire naturally small
Dirac masses if the left-handed neutrinos on our wall couple to any
massless
bulk fermion. Since the SM gauge fields are localized on our 3-brane, a
bulk fermion
must be a SM singlet, and will henceforth be referred to as the
bulk right-handed neutrino in this section.
The reason for the suppressed mass is that bulk modes have couplings
suppressed
by the volume of the extra dimensions; this is the reason for the
weakness of
gravity at long distances in our scenario, as well as small gauge
couplings
for bulk gauge fields \cite{ADDlong,AHD,Ber}.

For simplicity, we begin by considering a toy 5 dimensional theory
to concretely illustrate the idea; the generalization to the physically
realistic case of
higher dimensions will then be clear. Consider a 5 dimensional theory
with
co-ordinates $(x^{\mu},y)$, with $\mu= 0, \cdots, 3$ and the $y$
direction
compactified on a circle of circumference $2 \pi R$ by making the
periodic
identification $y \sim y + 2 \pi R$. Our 3-brane, where the lepton
doublet and
the Higgs fields are localized, is located at $y=0$, while a
massless Dirac fermion
$\Psi$ propagates in the full five dimensions.
The $\Gamma$ matrices can be written as
\begin{equation}
\Gamma^{\mu} = \left( \begin{array}{cc} 0 & \sigma^{\mu} \\
\bar{\sigma}^{\mu}
& 0 \end{array} \right) , \, \Gamma^5 = \left( \begin{array}{cc} i & 0
\\
0 & -i \end{array} \right)
\end{equation}
where we have chosen the Weyl basis for the $\Gamma^{\mu}$ matrices.
The Dirac spinor $\Psi$ is also conveniently decomposed as usual in the
Weyl
basis
\begin{equation}
\Psi = \left(\begin{array}{c} \nu_R \\ \bar{\nu^c}_R \end{array} \right)
\end{equation}
Let us now shut off all interactions between bulk and wall fields and
understand the spectrum of the theory from the 4-dimensional point of
view.
If we Fourier expand
\begin{equation}
\nu_R^{(c)}(x,y) = \sum_n \frac{1}{\sqrt{2 \pi R}} \nu_{Rn}^{(c)}(x) e^{iny/R}
\end{equation}
then the free action for
$\Psi$ becomes
\begin{equation}
S^{\mbox{free}}_{\Psi} = \int d^4 x \sum_n \bar{\nu}_{Rn}
\bar{\sigma}^{\mu} \nu_{Rn} +
\bar{\nu^c}_{Rn} \bar{\sigma}^{\mu} \nu^c_{Rn} + \frac{n}{R} \nu_{Rn} \nu_{Rn}^c
+ \mbox{h.c.}
\end{equation}
Of course this is the usual Kaluza-Klein expansion, with the expected
result.
We have a tower of fermions $\nu_{Rn},\nu^c_{Rn}$ with Dirac masses $n/R$
quantized in units of $1/R$.
The free action for the Lepton doublet $l$ localized on the
wall is
just
\begin{equation}
S^{\mbox{free}}_{l} = \int d^4 x \bar{l} \bar{\sigma}^{\mu} l
\end{equation}
Let us now imagine writing down the most general interactions between
wall and
bulk fields. Since something analogous to Lepton number must be imposed
to
forbid too-large Majorana neutrino masses for the SM fields anyway, we
will
for simplicity assume that lepton number is
conserved and assign $\Psi$ has opposite lepton number as $L$.
The leading local interaction between $\Psi$ and wall fields is then
\begin{equation}
S^{\mbox{int}} = \int d^4 x \kappa l(x) h^*(x) \nu_R(x,y=0)
\label{intf}
\end{equation}
where $\ka$ is a dimensionless coefficient and we work in units where
the fundamental scale $M_* = 1$. Notice that this coupling manifestly
breaks the
full 5-dimensional Poincare invariance of the theory by picking out the
component $\nu_R$ from the full Dirac spinor $\Psi$. This is
perfectly reasonable, since the presence of the wall itself breaks the
5-dimensional
Poincare invariance to the 4-dimensional one, and therefore the
couplings
need only be invariant under the 4-d Poincare transformations.
As we show in
the appendix, this can be seen very explicitly in a specific set-up for
localizing
$l,H$ on a $3+1$dimensional domain wall in 4+1 dimensions.
Upon setting the Higgs to its vacuum expectation value $v$, and
expanding
$\psi(x,y=0)$ in KK modes, the above interaction generates
the following mass terms
\begin{equation}
S^{\mbox{int}} = \int d^4 x \frac{\kappa v}{\sqrt{2 \pi R}} \nu_L(x)
\sum_n \nu_{Rn}(x)
\end{equation}
Suppose that $\kappa v/\sqrt{R} \ll 1/R$. In this case, all the massive
KK excitations are unaffected by this term. However,
this interaction generates a Dirac mass term between $\nu_L$ and the
zero mode
$\nu_{R0}$, which is suppressed by the size of the dimensions:
\begin{equation}
m_{\nu} = \kappa \frac{v}{\sqrt{R}}
\label{na}
\end{equation}

It is clear that this generalizes to the case where the
right-handed neutrino lives in any number $n$ of extra
dimensions.
In the decomposition of a higher dimensional spinor
under the 4-d Lorentz group, there will be a number of left-handed Weyl
spinors
which can have an interaction of the type in Eqn.(\ref{intf}), which
gives a
mass term suppressed by (Volume$)^{-1/2}$ between $\nu_L$ and all the KK
excitations of the bulk right-handed neutrino. As long as this mass is
smaller than
$1/R$, this is negligible for the KK modes but gives a Dirac mass
\begin{equation}
m_{\nu} = \kappa \frac{v}{\sqrt{V_n M_*^n}}
\label{general}
\end{equation}
where we have restored the $M_*$ dependence.
Upon using the relation
$M_{pl}^2 = M_*^{n+2} V_n$, we obtain for the neutrino mass
\begin{equation}
m_{\nu} = \kappa \frac{v M_*}{M_{pl}} \sim 10^{-4} \mbox{eV} \times
\frac{
\kappa M_*}{1 \mbox{TeV}}
\end{equation}
Note that for all $n>2$, this mass is much smaller than $1/R$ so
our analysis was justified, while for $n=2$ they are roughly comparable;
this will pose phenomenological difficulties for $n=2$ as discussed in
section 5, and henceforth we shall only consider cases with $n>2$.
It is remarkable that for the case of a low string scale
$\kappa M_* \sim 1 - 100\tev$, this prediction for the neutrino masses
is very roughly in the right range to explain the atmospheric and solar neutrino anomalies.

Let us more carefully compute the neutrino mass, by integrating
out the KK modes. Integrating out the massive $\psi^{(\vec n)}, \psi^{c(\vec n)}$
pair at tree level generates the operator
\be
\frac{1}{|\vec n|^2/R^2} \bar{l} \bar{\sigma}^{\mu} \partial_{\mu} l h^* h
\ee
The sum over all KK modes is power divergent in the UV for $n>2$.
This UV divergence must be
cut-off
near the fundamental scale $M_*$, i.e. at a $|k|_{max}$ such that
$|k_{max}|/R = c M_*$, where $c$ is a dimensionless
factor reflecting our ignorance of where exactly this power
divergence is cutoff. The generated operator is
\begin{equation}
\frac{\kappa^2 c^{n-2}}{M_*^2} \bar{l} \bar{\sigma}^{\mu} \partial_{\mu}
l
h^* h
\label{nuLnorm}
\end{equation}
After setting the Higgs to its vev, this generates a correction to the
$\nu_L$ wavefunction renormalization. After going back to canonical
normalization
for $\nu_L$, the neutrino mass becomes
\begin{equation}
m_\nu = \frac{\kappa}{\sqrt{1 + \kappa^2 c^{n-2} v^2/M_*^2}} \times
\frac{v M_*}{M_{pl}}
\label{proper}
\end{equation}
The significance of this equation is that for a fixed value of $M_*$, it
is not
possible to increase the neutrino mass arbitrarily by increasing
$\kappa$,
rather there is an upper bound
\begin{equation}
m_{\nu}^{max} = c^{-(n-2)/2} M_*^2/M_{pl}
\end{equation}

All of this can be seen more explicitly by simply writing down
the mass matrix for the various neutrino fields; for
simplicity let us consider the case $n=1$. The relevant fields
with $L=1$ are $N_+=(\nu_L,\nu_{R1}^c,\nu_{R2}^c,\cdots)$, while
those with $L=-1$ are $N_-=(\nu_{R0},\nu_{R1},\nu_{R2}, \cdots)$. Note that $\nu_{R0}^c$
does not acquire a mass term with any other field and remains
exactly massless. The mass matrix is of the form
\be
{\cal L_{\mbox{mass}}} = N_-^T {\cal M} N_+,
\label{massterm}
\ee
with
\be
{\cal M} = \left(\begin{array}{cccc} m & 0 & 0 & \cdots \\ m & 1/R
& 0 & \cdots \\ m & 0 & 2/R & \cdots \\ \vdots & \vdots & \vdots &
\ddots \end{array} \right)
\label{massmatrix}
\ee
where $m$ is as in eqn.(\ref{na}).
If we treat all the off-diagonal terms as perturbations, then at
zero'th order the lightest eigenvalue of this matrix is $m$.
To first order in perturbation theory, the eigenvalues are
unchanged, but we find that the lightest $L=1$ mass eigenstate is
dominantly $\nu_L$, with an admixture of
\be
\theta_n \sim \frac{m}{|n|/R}
\ee
of the KK mode $\nu_{Rn}^c$.
The first shift in the eigenvalues occurs at second order in
perturbation theory. It is more convenient to use
the Hermitian matrix ${\cal M} {\cal M}^{\dagger}$, who's
eigenvalues are the absolute value squared of the eigenvalues of
$M$:
\be
{\cal M} {\cal M}^{\dagger} =
\mbox{diag}(m^2,(1/R)^2,(2/R)^2,\cdots) + \left(\begin{array}{cccc}
0 & 1 & 1 & \cdots \\ 1 & 1 & 1& \cdots \\ 1&1&1& \cdots \\
\vdots & \vdots & \vdots & \ddots
\end{array} \right) m^2
\ee
The lowest eigenvalue gets corrected to be
\be
m_{\nu}^2 \to m_{\nu}^2 \times \left( 1 - \sum_n \frac{m_\nu^2
R^2}{n^2} \right)
\ee
Taking the square root, this is nothing but the first term in the
perturbative expansion of eqn.(\ref{proper}).

\subsection{Right-handed neutrinos from sub-spaces}

The bulk fermion fields that give rise to the right-handed neutrinos on our
brane do not necessarily live in the entire transverse
$n$-dimensional bulk. It is consistent to suppose that they
propagate in just an $m$-dimensional subspace ($m<n$) of the
entire bulk where gravity propagates. Such a
situation can easily arise if our three-dimensional world
is at the intersection point of two or more branes with at least
one having $p=m+3>3$ spatial dimensions. Independent of how
such a scenario is realized, the properties of the right-handed neutrino
interactions with our wall localized states are simply described as a simple extension of
the discussion in the previous section, which we do in a slightly different way below.
Denote by $V_m$ the $m$-dimensional transverse volume in which the
right-handed neutrino field propagates. Then once again the KK mode
expansion of this field is
\be
\nu_R(x,y) = {1\over \sqrt{V_m}} \sum_{\vec \ell} \nu_{R,\vec \ell}(x)
\exp( -2\pi i {\vec \ell}\cdot {\vec y}/(V_m)^{1/m} ).
\label{mdimKK}
\ee
The interaction of the KK zero mode $\vec\ell =0$ with an operator
${\cal O}$ constructed out of wall-localized
standard model states is still given by the overlap integral
\be
{\rm Prob} = \int d^3 x d^n y {\cal O}_{\rm SM}(x) \nu_{R,0}(x,y).
\label{overlap}
\ee
Each standard model field in ${\cal O}$ has in it's wavefunction
a factor of $1/\sqrt{V_{\rm wall}}$ arising from the small wall extent
in the
$m$ transverse dimensions. Furthermore there is a factor of
$1/\sqrt{V_m}$ from the normalization of the right-handed neutrino state, and
a factor of $V_{\rm wall}\sim 1/\mst^m$ coming from the $\int d^n y$
integral
which is only non-zero in the $m$-dimensional subspace where both
the wall extends and the right-handed field propagates. Putting this together
in the case of interest, the interaction term $\nu_R L H$ is suppressed
by the probability
\be
{\rm Prob} = \left( {V_{\rm wall}\over V_m}\right)^{1/2}.
\label{prob}
\ee
In the case of a symmetric internal manifold where each of the
$n$ dimensions is of size $R$, the volume of the $m$-dimensional
subspace is $V_m \sim R^m$. Thus upon using $\mpl^2 = R^n \mst^{n+2}$
the factor in Eqn.~(\ref{prob}) reduces to
\be
{\rm Prob} = \left( \mst\over \mpl\right)^{m/n}.
\label{prob2}
\ee

Including the power divergence of the normalization of the $\nu_L$
kinetic term, Eqn.(\ref{nuLnorm}, adapted for the case where the
right-handed
neutrino propagates in $m<n$ dimensions, we have (for all the large
dimensions of roughly equal size) the neutrino mass expression
\begin{equation}
m_{\nu} = \frac{\kappa v}{\sqrt{1 + c^{m-2} v^2/M_*^2}}
\times \left(\frac{M_*}{M_{pl}}\right)^{m/n} .
\end{equation}
Thus a large spectrum of neutrino masses is possible.
For instance, if $n=6$ and
$m=5$, even for $\kappa \sim 1$ and $M_* \sim 1\tev$, we get
$m_{\nu} \sim 10^{-2}\ev$, naturally the correct order
of magnitude for explanations of the atmospheric neutrino anomaly.

In general we should note that there is no reason for the internal
$n$-dimensional
manifold to be symmetric. For instance in the case $n=6$ we could
imagine
compactifying on a product of two-tori $T^2 \times T^2 \times T^2$, each
with
its' own characteristic radius. The Gauss' law condition for $\mpl$
only
requires that the total volume $V_m = \mpl^2/\mst^{n+2}$. If we now
define
an average radius $R$ by the relation $R^n=V_n$, and write $V_m =
V_n/V_{n-m} =
R^n /V_{n-m}$, we get the general form of the suppression for the
coupling
$\nu_R L H$;
\be
{\rm Prob} = \left( \mst\over \mpl\right)^{m/n}
\left( {V_{n-m}\over R^{(n-m)}}\right)^{1/2}.
\label{probasymm}
\ee


\section{Breaking lepton number on distant walls}
In the previous sections, we have considered ways of obtaining
naturally small {\it Dirac} masses for the neutrinos, in theories
with conserved lepton number. It is also possible to generate
small {\it Majorana} neutrino masses, by using the generic
idea of \cite{AHD} for generating small couplings by breaking
symmetries on distant branes. In our case, we wish to imagine that
lepton number is primordially good on our brane, but is maximally
badly broken at the scale $M_*$ by the vev of a field $\phi_L$ with lepton number
$L=2 $ on a different brane located at $y=y_*$ in the extra dimensions.
The
information of this breaking is transmitted to us by a bulk field $\chi_L$
also carrying $L=2$. Working in units with $M_*=1$, the relevant
interactions are
\be
{\cal L} \supset \int_{\mbox{us}} d^4 x \kappa (l h^*)^2(x)
\chi_{L}(x,y=0) + \int_{\mbox{other}} d^4 x' \langle \phi_L
\rangle \chi_L(x,y=y_*)
\ee
The vev of $\phi_L$ on the other brane acts as a source for
$\chi_L$, and ``shines" $\chi_L$ everywhere. In particular, the
shined value of $\chi_L$ on our brane is just given by the Yukawa
potential  in the transverse $n$ dimensions \cite{AHD}
\be
\langle \chi\rangle (x,y=0)= \Delta_n(|y_*|), \Delta_n(r) =
\left(-\nabla^2_{(n)} + m^2_{\chi_L}\right)^{-1}(r).
\ee
For $n>2$,
\begin{eqnarray}
\Delta_n(r) &\sim& \frac{e^{-m r}}{r^{n-2}} \, \mbox{for} \, m r \gg 1,
\nonumber \\
&\sim& \frac{1}{r^{n-2}} \, \mbox{for} \, m r \ll 1
\end{eqnarray}
The resulting Majorana neutrino mass is suppressed by the factor
$\Delta_n(|y_*|)$, restoring the dependence
on $M_*$ we have
\be
m_{\nu}^{\mbox{Maj.}} \sim \frac{v^2 \Delta_n(r)}{M_*^{n-1}}
\ee
This can give us an exponential suppression if $\chi_L$ is massive,
while even if $\chi$ is very light, a power suppression is
possible.

The case of massive $\chi_L$ can easily generate small enough
Majorana masses, but is not particularly predictive without a
theory specifying the inter-brane potential. On the other hand, if
we consider very light $\chi_L$ (i.e. lighter than $1/R$ but heavier
than $\sim ($mm$)^{-1}$ to have escaped detection), and assume
that the brane where $L$ is broken is as far away as possible i.e.
that $|y_*| \sim R$, then the neutrino mass is predicted to be
\be
m_{\nu}^{\mbox{Maj.}} \sim \frac{v^2}{M_*} \left(\frac{M_*}{M_{\rm
pl}}\right)^{2 - 4/n}
\label{light}
\ee
where we have used $M_*^{n+2} R^n \sim M_{\rm pl}^2$.
Note that for $n=4$, we recover the same rough prediction for
neutrino masses as the old see-saw mechanism and the bulk
right-handed neutrino. In this case there is a little more
flexibility since the walls don't have to be quite so far away,
and this can enhance the neutrino mass in the correct direction.

\section{Neutrino masses from the brane-lattice crystal}
A qualitatively different possibility is raised if we are willing
to contemplate a bulk populated with large numbers of branes. This
possibility was raised in \cite{AHDMR} in the context of
stabilizing the extra dimensions; where the largeness of the extra
dimensions was linked to the large brane number. For our purposes
here we simply assume that the bulk is populated with a number density
$n_{brane}$ of branes. In order to have a consistent picture of
the brane lattice ignoring quantum gravitational effects, we must
require that the lattice is dilute on the fundamental Planck
scale i.e.
\be
n_{brane} \ll M_*^n
\ee
Let us assume that lepton number is broken
on about  half of the branes, while it is unbroken on the other
half; our brane happens to be one where $L$ is unbroken. The
information of $L$ breaking is transmitted to us by bulk
messengers $\chi_L$ as in the previous section, leading to a
neutrino mass
\be
m_{\nu}^{Maj.} \sim \frac{v^2}{M_*^{n-1}} \int d^n y \, n_{brane}
\Delta_n(|y|)\nonumber
\ee
Let us now suppose that $\chi_L$ is massive enough so that its
Compton wavelength is smaller than the distance to the nearest
wall. Then,
\begin{eqnarray}
m_{\nu}^{Maj.} &\sim& \frac{v^2}{M_*^{n-1}} \int dr r^{n-1} \, n_{brane}
\frac{e^{-m_{\chi_L} r}}{r^{n-2}} \nonumber \\
& \sim & \frac{v^2 n_{brane}}{M_*^{n-1} m_{\chi_L}^2}
\end{eqnarray}
It is perhaps most natural in this context to take $m_{\chi_L} \sim
M_*$, in which case the smallness of the neutrino mass is wholly
controlled by the brane density. In the brane-lattice
crystallization scenario for radius stabilization, this density
was determined to be \cite{AHDMR}
\be
\frac{n_{brane}}{M_*^n} \sim \left(\frac{M_*}{M_{\rm
pl}}\right)^{4/n}.
\ee
Using this value for the density leads to a neutrino mass
\be
m_{\nu}^{\mbox{Maj}.} \sim \frac{v^2}{M_*} \left(\frac{M_*}{M_{\rm
pl}}\right)^{4/n}.
\ee
Again the case $n=4$ leads to a neutrino mass of roughly
the correct order of magnitude for solar and atmospheric
neutrinos, with $n_{brane}$ and $m_{\chi_L}$ varying over
reasonable ranges.

\section{Phenomenological constraints}
The main constraints on any theory with SM fields localized on a
3-brane have to do with production of light bulk modes. The
graviton is the one model-independent example of such a field, and
graviton overproduction in astrophysical systems
and in the early universe place unavoidable constraints on our
framework, but do not exclude it \cite{ADDlong}. As discussed in
\cite{AHD}, if there are other light states in the bulk, such as
vectors and scalars, even stronger bounds can result. The reason
can be understoof by simple dimensional analysis. The bulk
graviton couples to dimension 4 operators on the brane. As such,
working in terms of the canonically normalized bulk
graviton field $h_{AB}$, which has mass dimension $(n+2)/2$, the
coupling is schematically of the form
\be
\int d^4 x {\cal O}_4 (x) \frac{h}{M_*^{(n+2)/2}}
\ee
and therefore the cross sections for graviton emission scale with
the energy as
\be
\sigma(\mbox{grav. prod.}) \sim \frac{E^n}{M_*^{n+2}}
\ee
By contrast, a vector field in the bulk couples to a dimension 3
operator on the wall,
\be
\int d^4 x {\cal O}_3(x) \frac{A}{M_*^{n/2}}
\ee
and the rate for bulk vector production is correspondingly
enhanced
\be
\sigma(\mbox{vect. prod.}) \sim \frac{E^{n-2}}{M_*^n}
\ee
By this reasoning, the right handed neutrino, coupling as it does
to the lowest dimension SM invariant operator on our wall, should
be most strongly coupled and potentially dangerous. However, it is
important to remember that being a SM singlet, the bulk neutrino
only interacts with SM fields via its mixing to $\nu_L$.

First consider putting the Higgs to its vev (we will return to
processes involving physical Higgs fields at the end of this
section). Then, the coupling of the right-handed neutrino to the
wall neutrino generates a small Dirac mass as we have seen, with
the lightest state being predominantly $\nu_L$ but having an
admixture of the higher KK excitations of $\nu_{Rn}^c$. For $n=2$,
this mixing can be $O(1)$ and disastrous, while for higher $n$,
even though the mixing to each state is small, the large multiplicity of
states can still potentially give problems. It is most convenient
to first go to the mass eigenstate basis. Then, the tower of
$\nu_{R \vec n}^c$ KK states only interact through gauge interactions,
with the vertices suppressed by $\theta_{\vec n} \sim m_{\nu}/(|\vec
n|/R)$. Let us consider the implications of this for early
universe (but post ``normalcy temperature" $T_*$ \cite{ADDlong})
cosmology.

First, we have to determine whether any of these KK modes are
ever thermalized. The worst case (biggest mixing angle) is for the
first KK mode. The thermalization proceeds through through $W,Z$
exchange with ordinary SM particles, with a cross section
\be
\sigma \sim G_F^2 T^2 \theta^2.
\ee
We determine the decoupling temperature as usual by equating $n \sigma v = H \sim T^2/M_{\rm
pl}$, which yields
\be
T_{\rm dec.} \sim 1 \mbox{MeV} \, \theta^{-2/3}.
\ee
For $n=2$, the situation is problematic, and likely too many of
the heavy modes will be thermal during nucleosynthesis. However,
already for $n=3$, the largest $\theta \sim 10^{-5}$ even taking $m_v \sim 3 \times
10^{-2}$eV for the atmospheric neutrino problem, and the
decoupling temperature is forced above $\sim 1$ GeV. Since in all
cases, the normalcy temperature $T_* \lsim 1$ GeV, we can conclude
that for $n>2$, the KK neutrinos are never thermalized once the
universe becomes ``normal". Of course, we have to insure that
they, and more importantly bulk gravitons, are not created in
thermal abundances {\it before} $T_*$, but that is a separate
issue of the very early universe cosmology in this scenario which
we will not address here.

Next, just like the non-thermalized bulk gravitons, there is the
worry of evaporating  too much energy into these bulk neutrino
modes, unacceptably altering the expansion rate of the universe.
First, we need to determine the rate at which any given KK mode of
mass $m_{KK}$ decays back into SM states. The width is given by
\begin{eqnarray}
\Gamma &\sim& G_F^2 m_{KK}^5 \left(\frac{m_{\nu}}{m_{KK}}\right)^2
\nonumber \\
\Gamma^{-1} &\sim& 10^7 \mbox{s} \frac{10^{-3}
\mbox{eV}^2}{m_{\nu}^2} \left(\frac{1 \mbox{GeV}}{m_{KK}}\right)^3.
\end{eqnarray}
Note that
the KK modes produced at temperatures beneath $\sim 1$ GeV
are still around during nucleosynthesis. The rate which energy is
evaporated into bulk neutrinos at temperature $T$ is
\be
\dot{\rho}_{\nu^c_{R}} \sim -\frac{T^{n+7} m_{\nu}^2 G_F^2 M_{\rm
pl}^2}{M_*^{n+2}}
\ee
to be compared with the normal cooling rate by adiabatic expansion
\be
\dot{\rho}_{normal} \sim - \frac{T^6}{M_{\rm pl}}.
\ee
Requiring the normal rate to dominate over the neutrino rate at
least for $T \sim MeV$ when nucleosynthesis happens puts a rather
mild bound on $M_*$,
\be
M_* > 10^{\frac{14 - 6n}{n+2}} \mbox{TeV}.
\ee
The reason for the weak bound is that production of bulk $\nu$
modes must proceed through a $W/Z$ and is therefore further
suppressed by a $G_F^2$ factor.
Of course we in principle have to worry about the decays of these
bulk modes. The bulk gravitons which are produced have a long
lifetime of order the age of the universe and can unacceptably
alter the background gamma ray spectrum when they decay. Bulk
neutrinos are not as long-lived,and can be made to decay more
harmlessly on a ``fat brane" \cite{ADDlong} just as in the case of bulk
gravitons. Furthermore, if they decay to relativistic matter on
the other brane, there is no worry that there decay products will
ever overclose the universe.
Other phenomenological constraints on right-handed neutrinos are
similarly safe, for the same reasons.

One place for interesting
signals could be in physical Higgs decays to $\nu_L + $ bulk
neutrino. The width for the decay into any KK mode is suppressed
by the neutrino Yukawa coupling $\lambda_\nu^2 = m^2_{\nu}/v^2$,
but there is an enhancement $\sim (m_H R)^n$ coming from the sum over
all KK modes. The total decay width is
\be
\Gamma_{H^0 \to \nu_L \nu_R} \sim \frac{m_H}{16 \pi} \times
10^{3-n} \times \left(\frac{m_{\nu}^2}{10^{-5} \mbox{eV}^2}\right)
\times \left(\frac{m_H}{100 \mbox{GeV}}\right)^n \times \left(\frac{1
\mbox{TeV}}{M_*}\right)^{n+2}.
\ee
This invisible decay for the Higgs has a significant rate for
$n=3$! A detailed analysis of novel Higgs physics, both in this
scenario for generating neutrino masses as well as in
extra-dimensional flavor theories will be reported elsewhere.

Finally, the constraints on light bulk $\chi_L$ messengers are
essentially the same as those studied in \cite{AHD}, and just as
the cases studied there, the exchange of the light $\chi$ field
can give rise to attractive, isotope dependent sub-millimeter
forces $\sim 10^6$ times stronger than gravity, a signal that can
not be missed by the upcoming generation of sub-mm gravitational
force experiments.

\section{Large neutrino magnetic moments}
As an example of other
interesting neutrino physics in our scenario, we comment that it
may be possible to generate large neutrino magnetic moments
without neutrino masses. Suppose that there is an $SU(2)$ symmetry
acting on the left handed doublets of the SM. Then, the $SU(2)$
invariant Majorana mass term $\nu_a \nu_b \epsilon^{ab}$ vanishes
by antisymmetry. On the other hand, a magnetic moment operator of
the form $\nu_a \sigma^{\mu \nu} F_{\mu \nu} \nu_b \epsilon^{ab}$
is not constrained to vanish. Note that this $SU(2)$ symmetry
must be broken in order to generate charged lepton mass
splittings. However, it is easy to arrange this while still
forbidding neutrino masses. For instance, suppose that the flavor
symmetry is $U(2)_L \times U(2)_R$ \cite{AHD,Ber}. If this symmetry is broken by
a bi-fundamental, then charged lepton masses can arise, while
Majorana neutrino masses are still forbidden.
Since the UV cutoff in our framework
is only $\sim$ TeV, we can have the magnetic moment operator
suppressed by $\sim$ TeV, generating a large neutrino magnetic moment
$\sim 10^{-19} e$ cm in the absence of a neutrino mass.

\section{Conclusions}
Theories which lower the fundamental scale of gravity close to TeV
energies do not allow for the large desert in energy space between
$\sim 10^3 - 10^{19}$ GeV which have previously proven useful in
model-building. In particular, we seem to lose the
see-saw mechanism for explaining small neutrino masses, since the requisite large energy scale for
the right-handed neutrino mass is no longer at our disposal.
In this letter, we have shown that there are instead new, {\it intrinsically higher-dimensional}
mechanisms for generating small neutrino masses. We explored two options.
The first mechanism identifies  right-handed neutrinos with any
massless bulk fermions. These have volume suppressed
couplings to the left-handed neutrino localized on our
three-brane, and can generate naturally small Dirac neutrino
masses. The second mechanism takes advantage of the large {\it
infrared} desert in our scenario: the large space in the extra
dimensions. As an application of the general mechanism of
\cite{AHD}, small Majorana neutrino masses can result if lepton
number is broken on distant branes, with the breaking being
communicated to our wall by bulk messengers.
In this letter we have been content to show that the neutrino
mass scales required for explaining the atmospheric and solar
neutrino problems can naturally arise in our framework, while we
have left the flavor structure unspecified. Of course these could
come about in a fairly standard way through flavor symmetries,
although intrinsically higher-dimensional
scenarios would be more interesting.
We expect that in this and other areas,
model-building in extra dimensions will continue to be rich with fresh
possibilities for phenomenology.

\section*{Acknowledgments}
It is a pleasure to thank G. Farrar, L.J. Hall, A. Smirnov for
valuable discussions.
SD thanks the CERN theory group, and JMR thanks the Stanford University
theory group for their respective hospitality during portions of this
work.
The work of NAH is supported by the Department of Energy under contract
DE-AC03-76SF00515. The work of SD is supported in part
by NSF grant PHY-9870115. The work of JMR is
supported in part by an A.P.~Sloan Foundation Fellowship.

Note added: Yesterday, we received a
paper by Dienes, Dudas and Gheghetta \cite{DDGnu} which considers
a different mechanism: the possibility of neutrino oscillations without neutrino masses.
We do not believe that their mechanism works, however, since
motion in extra dimensions can not change lepton number.
Specifically, their analysis is based on a mass matrix (their eqns. (2.9),(2.10))
where the KK modes have {\it lepton number violating} masses.  However, these
masses, coming from the kinetic term in higher dimensions, must
{\it conserve} lepton number (see our e.g. eqn.(\ref{massterm}) and mass matrix
(\ref{massmatrix})).

\section*{Appendix}
In this appendix we wish to show more explicitly that an interaction of
the
form of Eqn.(\ref{int}), which is manifestly non-invariant under 5
dimensional
Poincare invariance, can nevertheless be generated in a theory
where the 5-d Poincare invariance is spontaneously broken by the
domain wall on which $l,H$ are localized.
Let $\Phi_W$ be a real scalar field whose vev breaks some discrete $Z_2$
symmetry,
the ``kink" configuration interpolating between two vacua
\begin{equation}
\langle \Phi_W(y \to \infty) \rangle = + \Phi_\infty, \,
\langle \Phi_W(y \to -\infty) \rangle = -\Phi_\infty, \,
\end{equation}
gives rise to a domain wall. The position $y_{wall}$ of the wall in the
fifth
direction is arbitrary, so translations in this direction are
spontaneously
broken. The associated Nambu-Goldstone $g(x)$ just corresponds to the
sound
waves on the wall, that is to the deformations
\begin{equation}
\Phi_W(x,y) = \langle \Phi_W(y + g(x)) \rangle
\end{equation}
Following the same sorts of arguments as in \cite{ADD}, we can easily
trap
chiral fermions ($l$ in this case) and scalars ($h$) on the domain wall.

Let us recall how $l$ can be trapped. Introduce a 5-dimensional Dirac
spinor
\begin{equation}
L = \left(\begin{array}{c} l \\ \bar{l}^c \end{array} \right)
\end{equation}
which has a Yukawa coupling to the wall field
\begin{equation}
\int d^4x dy \Phi_W \bar{L} L
\end{equation}
It is then well-known that zero modes of the Dirac equation
in the wall background exist of the form
\begin{equation}
L = \left(\begin{array}{c} f(y) l \\ 0 \end{array} \right)
\end{equation}
where $f(y)$ is normalizable i.e. $\int dy |f|^2 = 1$, whereas solutions
of the
form
\begin{equation}
L = \left(\begin{array}{c} 0 \\ g(y) \bar{l}^c \end{array} \right)
\end{equation}
are not normalizable $\int dy |g|^2 \to \infty$. Therefore, $l$ but not
$l^c$
is trapped to the wall. At distances large compared to the width of the
wall,
we can well approximate $f(y) = \sqrt{\delta(y)}$, and the localized
zero mode
is given by
\begin{equation}
L(x,y) = \left(\begin{array}{c} \sqrt{\delta(y)} l \\ 0 \end{array}
\right)
\label{trap1}
\end{equation}
Notice that the dimensionalities match: $L$ is a 5-d spinor
of mass dimension $2$, while $l$ is a 4-d spinor of mass dimension
$3/2$, the
difference being made up by $\sqrt{\delta(y)}$ which has mass dimension
$1/2$.
Similarly, it is easy to trap scalar field $h$ on the wall from a bulk
scalar
field $H$ coupled to the wall field (for more details see \cite{ADD}).
Again, at long distances the localized mode is
given as
\begin{equation}
H(x,y) = \sqrt{\delta(y)} h(x)
\label{trap2}
\end{equation}
once again note that the mass dimensions match.
Now, suppose that the theory also had the $\Psi$ Dirac fermion (not
coupled to
the wall field), which coupled to $H$ and $L$ via
\begin{equation}
S^{\mbox{int}} = \int d^4x dy \kappa H^*(x,y) \bar{\Psi}(x,y) L(x,y)
\label{int}
\end{equation}
This gives some coupling between the trapped modes on the wall and
$\Psi$,
which can be read off by inserting Eqns.(\ref{trap1},\ref{trap2}) into
Eqn.(\ref{int})
\begin{eqnarray}
S &=& \int d^4 x dy \kappa \left(\sqrt{\delta(y)} h^*(x) \right)
\left( \sqrt{\delta(y)} l(x) \right) \nu_R(x,y)
\nonumber \\
&=& \int d^4 x \kappa h^*(x) l(x) \nu_R(x,y=0)
\end{eqnarray}
which is precisely the form of the interaction used in the main text.

\def\pl#1#2#3{{\it Phys. Lett. }{\bf B#1~}(19#2)~#3}
\def\zp#1#2#3{{\it Z. Phys. }{\bf C#1~}(19#2)~#3}
\def\prl#1#2#3{{\it Phys. Rev. Lett. }{\bf #1~}(19#2)~#3}
\def\rmp#1#2#3{{\it Rev. Mod. Phys. }{\bf #1~}(19#2)~#3}
\def\prep#1#2#3{{\it Phys. Rep. }{\bf #1~}(19#2)~#3}
\def\pr#1#2#3{{\it Phys. Rev. }{\bf D#1~}(19#2)~#3}
\def\np#1#2#3{{\it Nucl. Phys. }{\bf B#1~}(19#2)~#3}
\def\mpl#1#2#3{{\it Mod. Phys. Lett. }{\bf #1~}(19#2)~#3}
\def\arnps#1#2#3{{\it Annu. Rev. Nucl. Part. Sci. }{\bf #1~}(19#2)~#3}
\def\sjnp#1#2#3{{\it Sov. J. Nucl. Phys. }{\bf #1~}(19#2)~#3}
\def\jetp#1#2#3{{\it JETP Lett. }{\bf #1~}(19#2)~#3}
\def\app#1#2#3{{\it Acta Phys. Polon. }{\bf #1~}(19#2)~#3}
\def\rnc#1#2#3{{\it Riv. Nuovo Cim. }{\bf #1~}(19#2)~#3}
\def\ap#1#2#3{{\it Ann. Phys. }{\bf #1~}(19#2)~#3}
\def\ptp#1#2#3{{\it Prog. Theor. Phys. }{\bf #1~}(19#2)~#3}

  \end{document}